\documentclass[aps,prl,twocolumn,groupedaddress,showpacs]{revtex4}

\usepackage{graphicx,amsmath,amssymb}
\usepackage{upgreek,xcolor}

\renewcommand{\vec}[1]{\mathbf #1}
\newcommand{\dd}{\mbox d}
\begin{document}


\title{Tube Width Fluctuations in F-Actin Solutions}

\author{J.~Glaser} 
\email{jens.glaser@itp.uni-leipzig.de}
\affiliation{Institut f\"ur Theoretische Physik,
  Universit\"at Leipzig, PF 100920, 04009 Leipzig, Germany}

\author{D.~Chakraborty}
\affiliation{Institut f\"ur Theoretische Physik, 
Universit\"at Leipzig, PF 100920, 04009 Leipzig, Germany}

\author{I.~Lauter}
\affiliation{Institut f\"ur Bio-~und Nanosysteme:~Biomechanik~(IBN4), Forschungszentrum J\"ulich, 52425 J\"ulich, Germany}

\author{M.~Degawa}
\altaffiliation[Current address: ]{Laboratory of Molecular Biophysics,
  Brain Science Institute, RIKEN -- Wako, Saitama 2-1, Japan}

\author{N.~Kirchge\ss ner}
\author{B.~Hoffmann}
\author{R.~Merkel}
\author{M.~Giesen}
\affiliation{Institut f\"ur Bio-~und Nanosysteme:~Biomechanik~(IBN4), Forschungszentrum J\"ulich, 52425 J\"ulich, Germany}
\author{K. Kroy}
\affiliation{Institut f\"ur Theoretische Physik, 
Universit\"at Leipzig, PF 100920, 04009 Leipzig, Germany}


\date{\today}

\begin{abstract}
We determine the statistics of the local tube width in F-actin solutions,
beyond the usually reported mean value.
Our experimental observations are explained by a segment fluid theory
based on the binary collision approximation (BCA). In this systematic
generalization of the standard mean-field approach effective polymer
segments interact via a potential representing the topological constraints.
The analytically predicted universal tube width distribution 
with a stretched tail is in good agreement with the data.

\end{abstract}

\maketitle

The Edwards tube model provides a simple phenomenological description of
the complicated topological constraints in entangled solutions of
flexible polymers \cite{Edwards1967}.  Using scaling arguments, Odijk
\cite{Odijk1983}, Semenov \cite{Semenov1986} and others have
generalized the idea to stiff polymers
with a persistence length
$\ell_p$ larger than the characteristic backbone length between mutual
collisions, which plays the role of the entanglement length
in this context.  Stiff polymers constitute an important and biologically
relevant class of polymers, as they represent the major structural and
stress-bearing units of the cytoskeleton of animal cells in the form
of filamentous actin and microtubules \cite{Kasza2007}.  Their large
$\ell_p$ and contour length $L$ (on the order of $10\,\upmu\mbox{m}$
for actin) have opened the possibility of direct microscopic
visualizations of the tube \cite{Kas1994}.
\begin{figure}[!ht]
\includegraphics[width=\columnwidth]{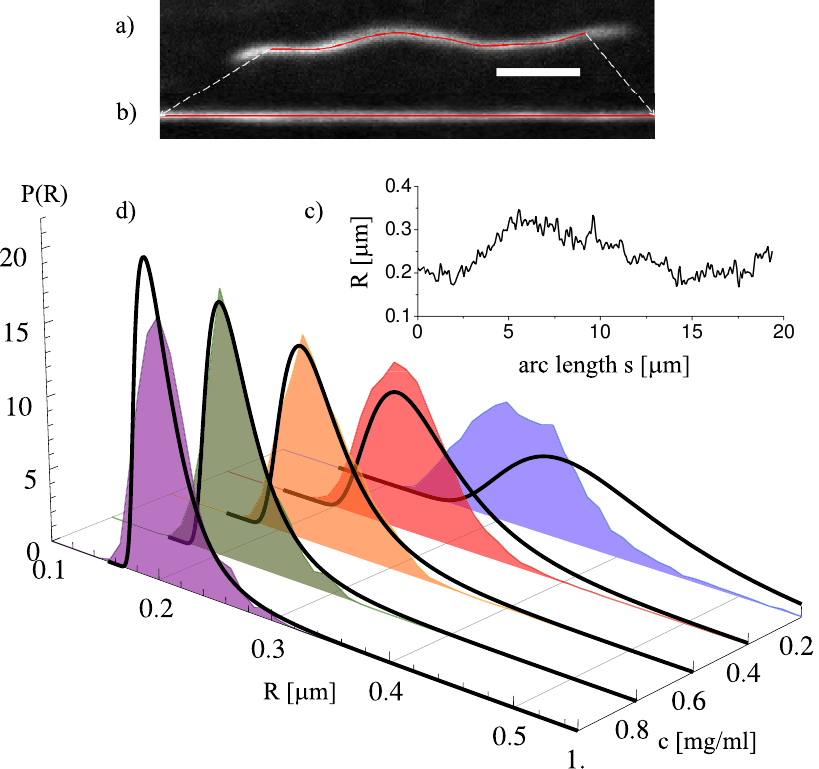}
\caption{a) Superimposed confocal microscopy images 
  of a fluorescent actin filament in a background solution and a spline
  representing the tube backbone; scale bar: $5\,\upmu\mbox{m;}$
  b)~rectified image; 
  c) tube radius profile $R(s)$ determined as standard deviation from
  Gaussian fits to the transverse intensity profile; d) normalized
  tube radius distribution $P(R)$ obtained from cumulative contour
  lengths of $536$, $804$, $301$, $225$ and $116\,\upmu\mbox{m}$ for
  actin concentrations $c=0.2$, $0.4$, $0.6$, $0.8$,
  $1.0\,\mbox{mg/ml}$, respectively.  Solid lines represent a global
  fit by Eq.~\eqref{prgamma} with $\rho=5.95\, \upmu\mbox{m}^{-2}
  c[\mbox{mg/ml}]$ and $L=0.91 \times L_e^\infty$,
  including corrections accounting for the line spread function \cite{supplement}. }
\label{fig:experiment}
\end{figure}
Beyond its intuitive appeal and obvious relevance to single filament
experiments in entangled solutions \cite{Dichtl1999,Romanowska2009},
the tube enters explicit calculations of the overall mechanical
properties of stiff polymer solutions \cite{Morse2001,
  Hinsch2007,Fernandez2009}. The latter have been measured
rheometrically for biopolymers such as actin
\cite{Hinner1998,Gardel2003,Liu2006,Tassieri2008} and pectin
\cite{Vincent2007}, as well as for self-assembling synthetic networks
\cite{Wilkins2009,Kuhne2009}. While theoretical approaches so far
have employed a homogeneity (or mean-field) assumption, treating the
tube radius as a constant, experiments
\cite{Kas1994,Dichtl1999,Luan2008,Romanowska2009,Wang2010} indicate substantial
heterogeneities, which recently also have been found in simulations
\cite{Hinsch2007}. 

\begin{figure}
\includegraphics[width=\columnwidth]{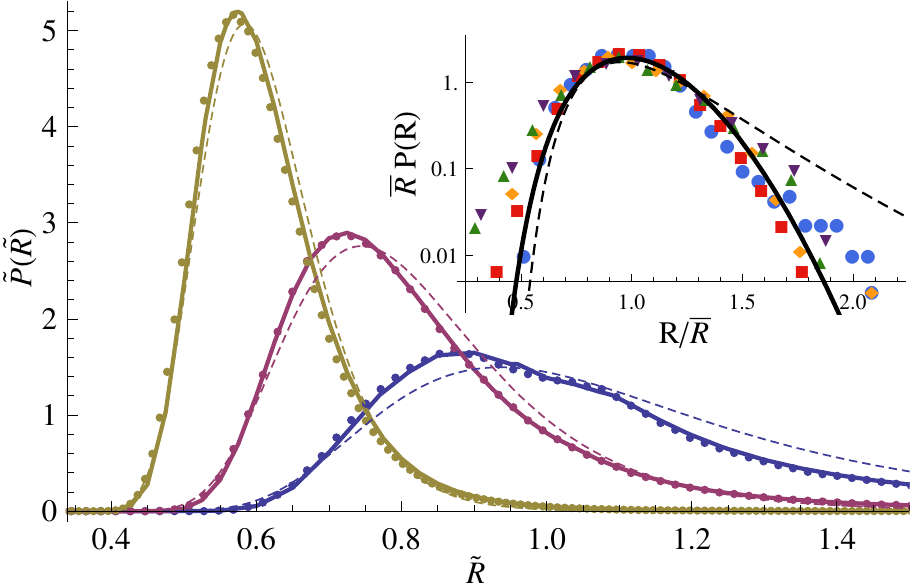}
\caption{Dimensionless tube radius distribution $\tilde P(\tilde R)$
  ($\tilde P\, \mbox{d}\tilde R = P\, \mbox{d}R)$
  with scaling variable
  $\tilde R \equiv L^{-3/8} {\overline R}^{-3/4} \ell_p^{1/8} R$ for
  reduced concentrations $\rho L \overline{R}=0.5,1,2$ (bottom to top):
  Holtzmark distribution, Eq.~\eqref{holtzmark} (solid); Gamma distribution
  approximation, Eq.~\eqref{prgamma} (dashed);
  numerical integration (dotted). A bimodal structure develops if the
  theory is pushed towards the (unphysical) limit $\rho L \overline{R}\to 0$.
  {\em Inset}: Rescaled distribution
  as a function of $R/\overline{R}$ in semi-logarithmic representation:
  experimental data for $c=0.2$ (circles),
  $0.4$ (squares), $0.6$ (diamonds), $0.8$ (upright triangles) and
  $1.0\,\mbox{mg/ml}$ (downward-facing triangles), analytical approximation
  from Eq.~\eqref{prgamma} as in Fig.~\ref{fig:experiment} (dashed line),
  and self-consistent theory described in the main text with
  a slightly renormalized value of the segment length $L=1.62\times
  L_e$ (solid line). }
\label{fig:prsim}
\end{figure}

\begin{figure}
\includegraphics[width=\columnwidth]{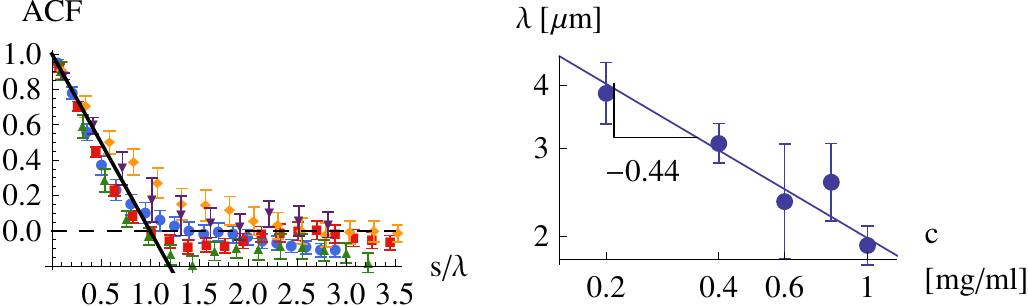}
\caption{ {\em Left:} Autocovariance function (ACF) of the tube radius profile
  averaged over $24$, $37$, $18$, $13$ and $9$ filaments for $c=0.2$,
  $0.4$, $0.6$, $0.8$, and $1.0\,\mbox{mg/ml}$,
  respectively (symbols as in Fig.~\ref{fig:prsim}, inset), with the
  abscissa rescaled by their initial slope $\lambda^{-1}$
  (obtained by quadratic extrapolation for a varying fit
  interval $s=s_0 \dots s_{max}$, $s_{max} \to 0$ \cite{supplement}).
  Because of the finite optical resolution of the microscope,
  data points below $s_0= 230\,\mbox{nm}$ were not considered.
  {\em Right}: $\lambda$ versus actin concentration $c$ with
  best power-law fit (exponent $-0.44 \pm 0.09$).}
\label{acf}
\end{figure}

In this Letter, we present a systematic study of these
heterogeneities in F-actin solutions, which we quantify in terms of the 
local tube radius profile $R(s)$ (cf.~Fig.~\ref{fig:experiment}a-c). From
$R(s)$ we infer the entanglement length (Fig.~\ref{acf}) and,
independently, the tube radius distribution $P(R)$ (Fig.~\ref{fig:experiment}d).
The latter plays a similar role as the
distribution of void spaces or pores in other disordered materials, such as
packings of grains or colloidal particles \cite{Torquato2002}.
As a main result, $P(R)$ is found to be described by a universal master function
with a stretched Gaussian tail (Fig.~\ref{fig:prsim}, inset).
We develop a systematic theory of tube fluctuations that explains our
data and provides the basis for a more comprehensive 
characterization of the structural and elastic properties
of stiff polymer solutions than possible within the
conventional mean-field theory \cite{Morse2001}. 

F-actin solutions were prepared at various monomer
concentrations $c$ \cite{supplement}.  Rhodamine-phalloidin labeled solutions
were mixed with
unlabeled solutions at a ratio of 1:1000. Time series of typically 150
pictures of individual labeled filaments were recorded using an inverse
confocal microscope (LSM510, Carl Zeiss Jena, Germany). These pictures
were superimposed to obtain a time-averaged image of the fluctuating
test filament, where intensity reflects the residence time of the filament
(Fig.~\ref{fig:experiment}a). Smooth tube contours 
connecting points of maximum intensity were constructed 
subject to a curvature-minimization constraint.
The {\em local tube 
radius} $R(s)$ was identified with the standard deviation of Gaussians
fitted to the transverse intensity profiles of a rectified tube image, as
exemplified in Fig.~\ref{fig:experiment}b $\&$ c \cite{supplement}.
Along a single test filament, $R(s)$ exhibits pronounced
fluctuations, which were binned to sample the tube radius distribution $P(R)$.
The result is depicted in Fig.~\ref{fig:experiment}d (shaded areas) for
various actin concentrations $c$.
The peak position and width, corresponding to the 
typical value and the fluctuations of $R$, respectively, are seen to decrease
with increasing $c$.  Yet, as we demonstrate below, the non-Gaussian
skewed and leptokurtic shape of $P(R)$ is well described by a $c$-independent
master function (Fig.~\ref{fig:prsim}, inset).

\begin{figure}
\includegraphics[width=\columnwidth]{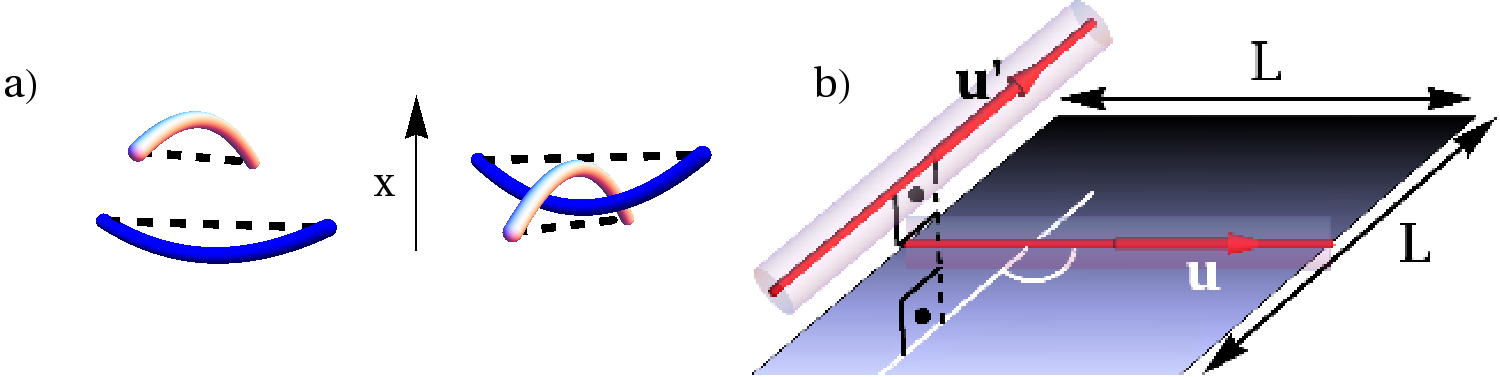}
\caption{a) A pair of colliding filaments is assigned a ``distance''
  $x$ and ``topology'' $\sigma=\pm$ as exemplified for $\sigma= +$
  with $x>0$ (left) and $x<0$ (right). A reflection at the horizontal
  mid-plane amounts to $x \to -x$, $\sigma \to -\sigma$. b) Overlap
  region (shaded) for two segments of length $L$ with
  orientations $\vec u$, $\vec u'$ enclosing an arbitrary angle. }
\label{overlap}
\end{figure}
We develop our theoretical approach along the lines of the BCA
\cite{Morse2001,Hinsch2007}. It replaces the complicated topological
constraints in an entangled polymer solution by an effective model:
an individual test polymer of length $L$ in a tube-like harmonic
confinement potential $\overline \phi \,\vec h^2\!/2$ per unit length,
where $\vec h(s)$ parametrizes the transverse contour undulations along the
backbone. In a self-consistent pair approximation, the tube potential
is required to represent the cumulative effect of independent pair
collisions with the background polymers. These collisions are not to
be understood as bare microscopic encounters of polymer backbones,
though, but rather refer to effective tube collisions. The bare
interactions caused by the impenetrability of the polymer backbones are
coarse-grained over the fast local conformational fluctuations of two
colliding polymers confined to their tubes, which prescribe a
``quenched'' geometry for them. The corresponding averages in a given
tube configuration and over different tube configurations are
represented by brackets $\langle \cdots \rangle_{\overline\phi}$ and over-bars
$\overline{\cdots}$, respectively. The over-bar on the tube stiffness
$\overline \phi$ thus indicates that the latter represents the cumulative
contribution from collisions in all possible geometries and topologies
as opposed to collisions in a prescribed tube configuration.
A subtlety in counting states is
that the topology of stiff but bendable polymers, as opposed to rigid
rods, is not uniquely determined by the center-of-mass positions and
orientations. Conversely, in any given topology $\sigma=\pm$ (or
``above'' and ``below''), positive and negative ``distances'' $x\lessgtr
0$ need to be distinguished, as sketched for $\sigma=+$ in
Fig.~\ref{overlap}a.  

The wormlike chain model in the weakly-bending rod limit
with eigenmodes $\vec h(q)$ of the undulations $\vec h(s)$ is employed.
Equipartition yields $\langle \vec h(q) \cdot \vec h(q') \rangle_{\overline\phi} =
2 \delta(q+q')/(\ell_p q^4 + \overline\phi)$
for $L \gg L_e^\infty$, where
$L_e^\infty \equiv (\ell_p/\overline\phi)^{1/4}$ is the mean-field
entanglement length \cite{Morse2001}.
Natural units with $k_B T=1$, so that $\ell_p$ is synonymous with the bending
rigidity, are used throughout.
The mean-field (projected) tube radius then follows as
$R^\infty\equiv R(\overline\phi)$,
given by the function
\begin{equation}
  \label{eq:R_phi}
  R^2(\phi) \equiv \int \frac
 {\dd s}{2L} \langle \vec h^2
 \rangle_{\phi}
  = 2^{-3/2} \ell_p^{-1/4} \phi^{-3/4}
\end{equation} 
evaluated at $\phi=\overline\phi$.
Similarly, one gets the partition sum
for the Gaussian contour undulations $\vec h(s)$ \cite{Morse2001}.
Its negative logarithm is the  mentioned coarse-grained interaction or ``BCA
potential''
$F_\sigma(x)=-\ln\{\mbox{erfc}[-2^{-1/2} \sigma x/ (R_0^2+R_1^2)^{1/2}]/2\}$
between two tubes of radii $R_0$ and $R_1$
at separation $x$ along the direction of nearest approach.
If we allow for a uniform transverse displacement $h$ of the test tube at an
angle $\psi$ relative to the $x$-direction, which does not change the topology,
the contribution to the confinement strength resulting from collisions in
the prescribed tube configuration follows as $\phi =
L^{-1} \partial_h^2 F_{\sigma}( x - h \cos\psi)|_{h= 0}$. 
In mean-field approximation, setting $R_0=R_1 \equiv R^\infty$ and
performing the configurational average \cite{Morse2001} yields the 
tube strength $\overline \phi$ as a function of $R^\infty$
and the line concentration $\rho$ (polymer length per volume).
From Eq.~\eqref{eq:R_phi}, 
$R^\infty = 0.66 \times \rho^{-3/5} \ell_p^{-1/5}$ is finally self-consistently
determined \cite{supplement}.

The conventional BCA, as a mean-field theory, is
exclusively concerned with the average values $\overline\phi$ and 
$R^\infty$.
To get hold of the measured spatial tube width fluctuations, we introduce
a {\em canonical ensemble of $N+1$ independent entanglement
segments}
of length $L$ characterized by their individual fields $\phi_i$ and
corresponding tube radii $R_i$ before averaging over the segment
ensemble.  In a formal generalization of the BCA that we call 
segment fluid model, any overlapping pair in the ensemble interacts with the
BCA pair potential, now written
as $V_{\sigma}(x) \equiv \chi(\vec r,\vec u,\vec u') F_\sigma(x)$ with the
characteristic function $\chi$ of the overlap between two segments
with orientations $\vec u$, $\vec u'$ separated by $\vec r$. As
depicted in Fig.~\ref{overlap}b, the two segments
are said to ``collide'' or to ``overlap'' if the
projection of the center-of-mass of the segment with
orientation $\vec u'$ onto the $\vec u$-$\vec u'$ plane 
falls into the shaded area with edges of length $L$ and
the segment with orientation $\vec u$ at its center.  All pairs of segments
are assigned a binary topological state variable $\sigma_{ij}=\pm$.  The
confinement potential for a test segment with index 0 is then computed as the
cumulative effect from the collisions with all overlapping
segments. Explicitly, after averaging over the uniformly distributed angles
$\psi_{0j}$ of perturbing displacements of the test segment, its individual tube
strength is written as a local field $\phi_0=\sum_{i=1}^N \phi_{0i}$ with
$\phi_{0i}= (2 L)^{-1} V_{\sigma_{0i}}''(x_{0i})$.
The average tube strength $\overline\phi$
is then obtained by taking the configurational average, i.e.\ by
summing $\phi_0$, weighted by the Boltzmann factor for all pair
interactions, over all topologies $\sigma_{ij}$ and positions
and orientations of the segments.
 
We now apply this formalism to calculate the distribution $P(\phi)$ of
tube strengths.  For the average over topologies we exploit the
identity $e^{-V_{+}}+e^{-V_{-}}=1$ for mutually overlapping segments, such that
Boltzmann factors not involving the test
segment cancel out in the partition sum. This is a consequence of the
topological origin of the effective pair interactions.  Moreover, due
to the pair approximation, all interacting pairs are equivalent so
that only one representative collision needs to be considered explicitly.
In the thermodynamic limit $N\rightarrow \infty$, the characteristic function
$\tilde P(t) =\overline{e^{i t\phi}}$ of $P(\phi)$ then follows from
the identity $(1+x/N)^N\to e^x$ as
\begin{equation}
  \tilde P(t) = \exp\!\left[n \int\dd \vec r_1\,\frac{\dd \vec
    u_1}{4 \pi}\, \Bigl(\mbox{$\sum\limits_{\sigma}$} e^{i t \phi_{01} - 
        V_{\sigma}(x_{01})}-1\Bigr) \chi \right]
\label{holtzmark}
\end{equation}
with the segment concentration $n\equiv (N+1)/V=\rho/L$.
By a numerical inverse Fourier transform, $P(\phi)$ is obtained, which we
recognize as the Holtzmark local field distribution of uncorrelated particles
\cite{Simon1990}. The tube radius distribution $P(R)$  then follows from
$P(\phi)$ by applying Eq.~(\ref{eq:R_phi}). In Fig.~\ref{fig:prsim}
(solid lines) it is shown for the special case of a pair potential $V_\sigma$
with a common average range $R_i \equiv \overline{R}$ for all segments $i$.

To validate the result, numerical (Monte Carlo) integration
of a fluid of effective segments under the same assumptions as used in the
theory was performed \cite{supplement}, see Fig.~\ref{fig:prsim} (dots).
An examination of the numerical results establishes that the
underlying asymmetric distribution $P(\phi)$ is well approximated by a
Gamma distribution $\Gamma_{k,\theta}(\phi)$.
Its shape and scale parameters, $k=4.013 \times \rho L \overline{R}$
and $\theta=0.125 \times (L \overline{R}^{2})^{-1}$,
are determined by matching its first two cumulants with the
predictions from Eq.~\eqref{holtzmark}.
Via Eq.~\eqref{eq:R_phi}, an analytical expression
for $P(R)$ ensues,
\begin{equation}
P(R)=\frac{8}{3 R (k-1)!} \exp(-y)\,y^k,\quad y\equiv
\frac{1}{4 \ell_p^{1/3} R^{8/3} \theta}.
\label{prgamma}
\end{equation}
It is compared with our numerical results
for a fixed prescribed segment length $L$ and various reduced concentrations
$\rho L \overline{R}$ in Fig.~\ref{fig:prsim}.

What remains to be done, is to identify the physical meaning of the
effective segment length $L$. Qualitatively, one expects it to be equal to
the mean-field entanglement length $L_e^\infty=(\ell_p/\overline{\phi})^{1/4}$.
The latter may in turn be expressed in terms of the mean-field tube
radius $R^\infty$ and the line concentration $\rho$, respectively. Via
Eq.~\eqref{eq:R_phi}, $L_e^\infty=
\sqrt{2} (R^{\infty})^{2/3} \ell_p^{1/3} \propto \rho^{-2/5}$.
For the natural choice of $L=L_e^\infty$, the mean tube radius
$\overline{R}\equiv\overline{R(\phi)}$ follows from a numerical solution of the
implicit equation $\overline{R}=\int \!\dd R\, R\, P(R)$ as
$\overline{R} = 1.15 \times R^\infty$.
Its close match with the mean-field value $R^\infty=\lim_{L\to\infty}
\overline{R}$ suggests to parametrize the distribution
$P(R)$ by the mean-field value $R^\infty \propto \rho^{-3/5}$ for
$\overline{R}$, with negligible error.  The shape parameter $k$ thereupon
becomes independent of $\rho$ and only depends on $L/L_e^\infty$.
The corresponding predictions of Eq.~\eqref{prgamma}
compare favorably with the measured tube radius distribution,
as demonstrated in Fig.~\ref{fig:experiment}d.
In this comparison, $L/L_e^\infty$ and $\rho/c$
are used as global concentration-independent
fit parameters.  While the fit does indeed corroborate the expectation
$L/L_e^\infty\approx 1$, $\rho/c$
turns out to be about a factor of $6.6$ smaller than estimated
from the molecular weight and structure of monomeric actin
\cite{Morse2001}. This numerical discrepancy can be eliminated without
significant consequences for the quality of the fit and the value of
$L/L_e^\infty$ \cite{endnote:corrections}.

Note that Eq.~\eqref{prgamma} implies that the concentration dependence
enters the tube radius distribution only through the average
tube radius $\overline R$, such that $\overline R P(R)$ defines
a concentration-independent master function of $R/\overline R$.
The inset of Fig.~\ref{fig:prsim} (symbols) demonstrates that the data indeed
scale satisfactorily.
The semi-logarithmic representation reveals some systematic deviation
of Eq.~\eqref{prgamma} (dashed) from the data, however. This shortcoming
is due to the preaveraging
$R_i \equiv \overline{R}$ employed in the derivation of Eq.~\eqref{prgamma},
and can be overcome by evaluating our systematic theory more accurately. To
this end, the variable tube radius $R_0$ of the test segment in the BCA
potential is self-consistently identified with the argument $R$ of the
distribution $P(R)$, and only the radius $R_1$ of
the representative collision partner is preaveraged.
This amounts to replacing $\overline{R}$ in
Eq.~\eqref{prgamma} by $2^{-1/2} [R^2+(R^\infty)^2]^{1/2}$,
before substituting $\rho =(R^\infty/0.66)^{-5/3} \ell_p^{-1/3}$.
Moreover, the segment length $L$ is taken to be proportional to the
{\em local} entanglement length, $L_e = \sqrt{2} R^{2/3} \ell_p^{1/3}$.
The resulting normalized tube radius distribution (solid line
in the inset of Fig.~\ref{fig:prsim}) is in excellent agreement with our
experimental data.
It exhibits (up to a spurious logarithm) a tail
 $P(R) \propto \exp[-(R/\overline{R})^{5/3}]$,
similar to what has recently been proposed on empirical
grounds \cite{Wang2010}.
 
A further consistency check for the developed theory is provided by the
spatial autocorrelation of the tube radius profile in
Fig.~\ref{fig:experiment}c.  It should decay over a characteristic
length scale comparable to the entanglement length.
The rescaled autocovariance functions determined from a large number of tube
radius profiles for various concentrations are shown in Fig.~\ref{acf}
(left panel), rescaled with their initial slope $\lambda^{-1}$.
Indeed, the concentration scaling of
$\lambda$ thus obtained (right panel) compares favorably with the expectation
$\lambda\simeq L_e^\infty \propto c^{-2/5}$.

The fact that heterogeneities give rise to stretched tails in $P(R)$
underscores the importance of shifting the attention from characteristic
``numbers'' for $R$, $L_e$ etc. \cite{Tassieri2008} to their skewed
leptokurtic distributions.
As demonstrated above, these are readily accessible in our
BCA-based segment fluid model.
For instance, subtle non-steric corrections to the value of the tube
radius $R$, e.g. due to transient electrostatic
attraction mediated by divalent counterions \cite{He2007}, could systematically
be studied by means of $P(R)$ in the future.
Our combined experimental and theoretical results might also
hold the key to a microscopic explanation of the ubiquitously observed
broad distribution of microrheological plateau moduli \cite{Waigh2005,
Luan2008}.

\begin{acknowledgments}
  This work was supported by the Deutsche Forschungsgemeinschaft
  (DFG) through FOR 877 and the Leipzig School of Natural Sciences --
  Building with Molecules and Nano-objects. M.~D. acknowledges financial
  support by the Alexander-von-Humboldt foundation.
\end{acknowledgments}

\end{document}